\RequirePackage{fix-cm}
\documentclass[smallextended]{svjour3}       
\smartqed  
\usepackage{amssymb}
\usepackage{amsmath}
\usepackage[latin2]{inputenc}
\usepackage{graphics}
\usepackage{graphicx}

\newcommand{\bea}{\begin{eqnarray}}

\newcommand{\eea}{\end{eqnarray}}

\newcommand{\be}{\begin{equation}}

\newcommand{\ee}{\end{equation}}

\newcommand{\at}[1]{{\bf \textit{#1}}}

\newcommand{\rt}[1]{{}}

\begin{document}

\title{Non-Gaussian fixed points in  fermionic field theories\\
without auxiliary Bose-fields }



\author{A. Jakov\'ac \and
        A. Patk\'os \and
        P. P\'osfay
}


\institute{A. Jakov\'ac \at
  Institute of Physics, E\"otv\"os University\\
  H-1117, P\'azm\'any P\'eter s\'et\'any 1/A, Budapest, Hungary\\
  \email{jakovac@caesar.elte.hu}
           \and
            A. Patk\'os \at
  MTA-ELTE Biological and Statistical Physics Research Group\\ 
  H-1117, P\'azm\'any P\'eter s\'et\'any 1/A, Budapest, Hungary\\
  \email{patkos@galaxy.elte.hu}
  \and
  P. P\'osfay \at
  Institute of Physics, E\"otv\"os University\\
  H-1117, P\'azm\'any P\'eter s\'et\'any 1/A, Budapest, Hungary\\
  \email{posfay.peter@gmail.com}
}

\date{Received: date / Accepted: date}

\maketitle

\begin{abstract}
The functional equation governing the renormalization flow of fermionic field theories is investigated in $d$ dimensions without introducing auxiliary Bose-fields on the example of the Gross-Neveu and the Nambu--Jona-Lasinio model. The UV safe fixed points and the eigenvectors of the renormalization group equations linearized around them are found in the local potential approximation. The results are compared carefully with those obtained with partial bosonisation. The results do not receive any correction in the next-to-leading order approximation of the gradient expansion of the effective action.
\end{abstract}

\section{Introduction}
The Functional Renormalization Group (FRG) has developed into an
important investigation tool of the large distance behavior of
strongly interacting quantum field theories
\cite{wilson74,wegner73,polchinski84,wetterich91,morris94}. In
particular, the emergence of bound states/condensates can be studied
very efficiently by introducing the corresponding composite fields
with appropriate quantum numbers into the set of operators from which
the low-energy effective action builds up
\cite{jungnickel96,gies02}. Typically, these fields are introduced at
short distances without the proper kinetic term through some
appropriate "Dirac-$\delta$" functionals equating composites formed
from the "microscopic" fields with the "observable" fields. Formally,
this construction associates with the new fields vanishing
wave-function renormalization constants $Z_{composite}(k=\Lambda)=0$
at the defining ultraviolet scale $\Lambda$.

If there are dynamical objects in the corresponding channel, their
wave-function renormalization should grow away from zero when one
reaches the compositeness scale. Below this momentum scale  a local
effective field, not revealing any internal structure should represent
them, possessing its own kinetic term. This expectation was checked in
the auxiliary field formulation of the $O(N)$-model \cite{gies02},
where the composite field introduced at the "microscopic" scale via
Hubbard-Stratonovich transformation, that is with no kinetic term,
became at low scale a propagating dynamical degree of freedom on its
own. Similarly, large anomalous scaling corrections were shown in the
Yukawa-coupling of boson-fermion models when searching for new
non-Gaussian (interacting) fixed points \cite{braun11}.

Ultraviolet stable non-Gaussian fixed points provide promising
alternative for consistent UV-completion of perturbatively
non-renormalizable theories, like the quantized Einstein-Hilbert
gravitational theory \cite{reuter02}. An asymptotic safety scenario
could circumvent the triviality problem of the Higgs-sector of the
Standard Model \cite{gies09,gies13}. Such ideas were put forward first time for the UV-completion of quantum electrodynamics \cite{miransky80,bardeen86,kogut88}. Another actual issue of interest
is to restrict the mass spectra of excitations in effective models of
particle physics, like the Higgs-sector of the Standard Model
\cite{gies14a,gies14b}. The study of analogous questions in simpler
theories helps to develop appropriate methods of investigations.

A compelling example for the above scenario is represented by theories
with four-fermion coupling like the Gross-Neveu \cite{gross74} or the
Nambu--Jona-Lasinio \cite{nambu61} model. One might introduce into
these theories bosonic fields corresponding to certain fermionic
bilinears through the delta functions
$\delta(\Phi^{a...}_{\mu...}(x)-(\bar\psi(x)\Gamma^{a...}_{\mu...}\psi(x)))$,
where $\Gamma^{a...}_{\mu...}$ is a conveniently chosen matrix with a
set of internal ($a$) and Lorentz ($\mu$) indices. At the
"microscopic" scale $\Lambda$ these fields do not have any
dynamics. In successful searches for non-Gaussian fixed points
substantial running of the wave-function normalization of the
composites has been observed and exploited \cite{gies02,braun11}.

In a recent paper we have proposed a scheme where one can explore the
fixed point structure of fermionic models in the framework of the
Wetterich-equation without introducing auxiliary variables
\cite{jakovac13}. Such approach has been introduced earlier in
connection with the dynamical breakdown of chiral symmetry in gauged
Nambu--Jona-Lasinio models \cite{aoki99,aoki00,aoki12,aoki14}. Fermionic evolution equations were developed
for QCD by Meggiolaro and Wetterich \cite{meggiolaro01} truncated at
the 4-Fermi level. Various three-dimensional theories have been
investigated recently including into the effective action the full set
of 4-Fermi operators taking into account Fierz-relations among them
\cite{gies10,braun12}. There is continuous interest in fermionic
theories also in condensed matter physics, where the expansion of the
effective action in powers of Fermi-fields is usually also truncated
at the level of 4-Fermi interactions, sometimes including also
momentum dependent (non-local) 6-Fermi vertices \cite{metzner12}. Our paper
constructed a rather general framework for the application of the
Renormalization Group method to purely fermionic relativistic field
theories without this limitation. For the fermionic "potential" of the
Gross-Neveu and the chiral Nambu--Jona-Lasinio model fully explicit
evolution equations were constructed. We focus in the present paper on
mapping the fixed point pattern of these theories and compare our
results with those obtained with approaches employing partial
bosonisation.

The running of the couplings starts slightly below the compositeness
scale, where one can treat the composite objects discussed above as
elementary (pointlike) and introduce an effective potential depending on arbitrary
powers of the invariants directly formed from the fermion
background. At high enough momentum, above the compositeness scale
higher powers of the invariants should get smeared into nonlocal
combinations of the Fermi-fields, reconciling in this way the
Grassmannian nature of these variables with the arbitrary powers
apparently present in the coarse grained potential.

Although there exists a considerable number of works in the literature
dealing with the local fermionic potential approximation
\cite{jakovac13,aoki99,aoki00,aoki12,aoki14}, and in our earlier
publication \cite{jakovac13} we have already worked out a rather
general framework for the treatment of this formalism, it is worth to
describe in a mathematically more accurate and less intuitive way how
the Local Potential Approximation (LPA) is introduced in the fermionic
case and what kind of approximations lie in the background.

\begin{figure}[htbp] 
\label{Fig:diagrams}
\centerline{ 
\includegraphics[width=3.5cm]{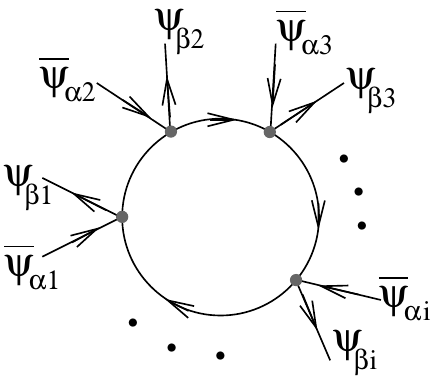}
}
\caption{One loop diagram contributing to the proper vertex
  $\Gamma^{(n)}_{k;\alpha_1\dots\alpha_n; \beta_1\dots\beta_n}(\{q\};\{p\})$. }
\label{one-loop-vertex}
\end{figure}

First of all we have to pin down that the fermionic effective action
contains an arbitrary power of the fermionic variables. Indeed,
$\Gamma_k[\bar\Psi,\Psi]$ is the generator of the proper multifermion
vertices at scale $k$:
\begin{equation}
  \Gamma[\bar\Psi,\Psi] = \sum_n \int dx_1dy_1\dots dx_ndy_n\,
  \Gamma^{(n)}_{k;\alpha_1\dots\alpha_n; \beta_1\dots\beta_n} (\{x\};
  \{y\}) \bar\Psi_{\alpha_1}(x_1) \Psi_{\beta_1}(y_1)\dots
  \bar\Psi_{\alpha_n}(x_n) \Psi_{\beta_n}(y_n).
\label{proper-vertex}
\end{equation}
These proper vertices are not zero for any $n$. An example of the
one-loop contribution in a theory with 4-fermion vertices (like the
Gross-Neveu or the Nambu--Jona-Lasinio model) can be seen in
Fig.~\ref{Fig:diagrams}.

\begin{figure}[htbp]
  \centering
  \includegraphics[width=4.2cm]{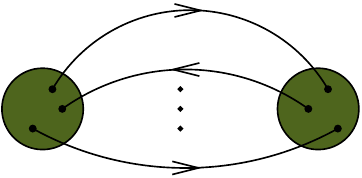}\hspace*{5em}
  \includegraphics[width=3.8cm]{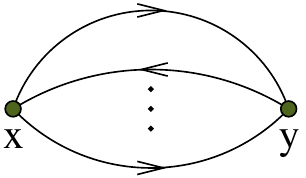}\\
  \hspace*{0.5em}\emph{a}\hspace*{16em}\emph{b}
  \caption{\emph{a}: A diagram generated by the proper vertices at
    scale $k$ and \emph{b}: its local approximation.}
  \label{fig:shrink}
\end{figure}

The assumption behind the LPA, both for bosons and fermions, is that,
for the running of the effective action, the most important
contributions come from that kinematical regime, where the vertex
varies much faster in spacetime than the propagators. This means that
any diagram contributing to the running which contains a nonlocal
proper vertex $\Gamma^{(n)}_k$ can be approximated by the contribution
where the vertex is concentrated to a single point,
cf. Fig.~\ref{fig:shrink}. The point-like vertex assumption is good if
the value of diagrams \emph{a} and \emph{b} on Fig.~\ref{fig:shrink}
are numerically close. Then we can replace the generator of the first
diagram which is the complete effective action, by the generator of
the vertices of the second diagram which is $\sum_{n>1} U_n
(\bar\psi(x)\psi(x))^n$. (Actually, also dependence on other fermion bilinears, compatible with Lorentz- and the internal invariance of the theory, is allowed. The restriction to the scalar combination does not restrict the generality of our arguments.)

When we seek a formal derivation, we make the approximation that the
value of the proper vertex $\Gamma_k^{(n)}(x_1,y_1,\dots x_n,y_n)$ is
zero (or sufficiently small) if $|x_i-x_j|>L$ or $|x_i-y_j|>L$, where
$L$ is the aforementioned compositeness, or localization length
scale. This means that all the fields are localized effectively within
a small neighborhood of $x_1$, we call it ${\cal V}_{x_1}$, its volume
we denote by $\Delta V$. According to the LPA we assume that the
fields are slowly varying on the scale $L$, then the proper vertex can
be substituted by an average value $\bar \Gamma^{(n)}_k$ (in
translation invariant systems it does not depend on the position).

The factorization of the average value out of the nonlocal vertex must be done
carefully, to avoid the appearance of a formal zero due to the
fermionic nature of the variables. The key 
observation is that
\begin{equation}
  \left(\frac1{\Delta V} \int_{{\cal V}_{x_1}}\!\!\! dx\, \bar\Psi(x)
    \Psi(x)\right)^{n}\neq0
\end{equation}
for any power $n$. Thus the approximate formula for the
effective action which corresponds to the diagram on the right of Fig. 2 is obtained by first putting the neighboring
$\bar\psi$ and $\psi$ fields to the same point, and then factoring out
the average value of the proper vertex. So we can write
\begin{eqnarray}
  &&\int dx_1dy_1\dots dx_n dy_n \Gamma^{(n)}_k (x_1,y_1,\dots x_n,y_n)
  \bar\Psi(x_1) \Psi(y_1)\dots 
  \bar\Psi(x_n) \Psi(y_n) \approx\nonumber\\&&\approx (\Delta V)^{n}
  \bar \Gamma^{(n)}_k\int dx_1 \int_{{\cal V}_{x_1}}\!\!\!\! dx_2\dots
  dx_n\bar\Psi(x_1) \Psi(x_1)\dots \bar\Psi(x_n) \Psi(x_n)
  =\nonumber\\&& = (\Delta V)^{2n-1} \bar\Gamma^{(n)}_k \int dx_1
  \bar\Psi(x_1) \Psi(x_1) \left(\frac1{\Delta V} \int_{{\cal V}_{x_1}}
    \!\!\! dx\, \bar\Psi(x) \Psi(x)\right)^{n-1}. 
\end{eqnarray}
As an effective way of writing we use the notation in the limit
$\Delta V\to0$:
\begin{equation}
   \Delta V^{2n-1} \bar\Gamma^{(n)}_k \int dx_1 \bar\Psi(x_1) \Psi(x_1)
  \left(\int_{{\cal V}_{x_1}}\!\!\! dx\, \bar\Psi(x)
    \Psi(x)\right)^{n-1} \!\!\! \stackrel{\Delta V\to
    0}\longrightarrow \;\;
  U_k^{(n)} \int dx (\bar\Psi(x) \Psi(x))^n. 
\end{equation}
This defines a (quasi) local potential for the fermionic fields. The
physical conclusion is that instead of point-like fermion fields we
should work with fermion bilinears averaged on patches, if we wish to study the pointlike limit of higher order 2n-fermion couplings.

The gradient expansion of the effective action is a series expansion
in increasingly non-local terms. It represents a unique hierarchy only
if the field content is not enlarged by introducing propagating
composite fields. Our aim is to study the gradient expansion in terms
of the original fermi-fields and compare with the results of a
different truncation of the derivative expansion arising from the
introduction of propagating auxiliary fields. Since the two cases have
different kinetic parts in the Lagrangian, one might expect different
convergence rate in the search for interacting UV-stable fixed point
theories.

In the present paper we shall demonstrate that some relevant results
demonstrated earlier in the auxiliary Bose-field formulation in the
next-to-leading order (NLO) of the gradient expansion of the
Wetterich-equation, can be obtained also in pure fermionic LPA,
provided we keep all powers in the fermionic potential. In particular, we study the effect of the
anomalous dimension of the wave-function normalization parameter of
the fermions and find that it vanishes in the ground state,
demonstrating this way the stability of our result at NLO of the
gradient expansion.

The paper is organized as follows. In section 2 we reformulate the
version of the Wetterich equation derived in \cite{jakovac13} for the
Gross-Neveu model in a space-dependent fermionic background. It is
projected on a constant background in Section 3, where the dependence
of the effective potential on a non-zero scalar composite condensate
is calculated. Here we present also the results of a fully analogous
study of the Nambu--Jona-Lasinio model. In Section 4 it is shown that
there is no anomalous scaling for fermionic fields, giving more
robustness to the results obtained in LPA. In the Conclusions we
compare our results with previous investigations. In particular, we
show that the $N_f=\infty$ results do reproduce all features of the
$d=3$ non-Gaussian asymptotically safe fixed point.
  
\section{Wetterich-equation for the $N_f$-flavor Gross-Neveu model in inhomogeneous background }

The Ansatz which corresponds to the next-to-leading order (NLO) of the gradient expansion of the Euclidean effective action $\Gamma$, taking into account the scale dependence of the wave function renormalization of the defining Fermi-fields is the following:
\be
\Gamma_k[\bar\psi,\psi]=\int_x\left[Z_k\bar\psi^\alpha_l(x)\partial_m\gamma_m^{\alpha\beta}\psi_l^\beta(x)+U_k(I(x))\right],\qquad I(x)=(\bar\psi\psi)^2\equiv (\bar\psi_l^\alpha(x)\psi_l^\alpha(x))^2,
\label{Gamma-ansatz}
\ee
where we have written out explicitly the bispinor index $\alpha$ and the flavor index $l$; the quantity $I$ (and with it the quantum action $\Gamma_k$) is invariant under the global discrete chiral symmetry transformation
\be
\psi\rightarrow -\gamma_5\psi,\qquad \bar\psi\rightarrow \bar\psi\gamma_5.
\ee

The operator content of the potential part of (\ref{Gamma-ansatz}) is
not complete. In principle all Lorentz and chiral invariant quartic
combinations should have been included. The number of independent
variables is then reduced by the Fierz-relations
\cite{jaeckel03,gies10,braun12}. Such an Ansatz can be called
Fierz-complete. It was established in \cite{jakovac13} that in the
complete expression of the right hand side of the Wetterich equation
only the invariant $I(x)$ appears, although in separate piecewise
contributions also other invariants (namely,
$(\bar\psi(x)\gamma_m\psi(x))^2$) are generated. This observation is
also true for the one-flavor Nambu--Jona-Lasinio model to be discussed
below. In view of these findings our study, though not being
Fierz-complete, is self-consistent.

It is worth to discuss in some detail the physical range of
variation of the invariant variable $I$ and the characteristics of the
potential $U$. This can be done explicitly for $N_f=\infty$ with help of the
auxiliary field formulation. First, we analyze the case
$U=g^2/(2N_f)I$. Its action is rewritten with the auxiliary field $\sigma(x)$ as
\be
\Gamma^{aux}_k[\bar\psi,\psi,\sigma]= \int_x
\left[Z_k\bar\psi^\alpha_l(x) \partial_m\gamma_m^{\alpha\beta}
  \psi_l^\beta(x)+\sigma(x)(\bar\psi\psi) -
  \frac{N_f}{g^2}\rho(x)\right],\qquad
\rho(x)=\frac{1}{2}\sigma^2(x).  
\ee
The model at $N_f=\infty$ is solved by finding the saddle point of
the effective action arising after integrating over the fermions
\cite{rosenstein89}. A phase transition occurs into the
broken symmetry phase at some $g_{cr}^2<0$. For $g^2<g^2_{cr}$ the
auxiliary field $\sigma$ has a non-zero, real expectation value $M$,
which determines also the size of the fermionic condensate: $\langle(\bar\psi\psi)\rangle=MN_f/g^2$. Since $g^2$ is negative, the
auxiliary potential $-N_f\rho/g^2>0$ can be interpreted as a
physically stable potential of the $\sigma$-field. This solution is
 matched with the mean-field potential of the original model by requiring
\be
\frac{g^2}{2N_f}I=M\langle(\bar\psi\psi)\rangle-\frac{N_f}{2g^2}M^2,
\ee
Substituting the saddle point value of $\langle(\bar\psi\psi)\rangle$, one recognizes that
$I$ varies along the positive axis, and its potential
energy is bounded from above in the broken symmetry phase.

This argument is made more general by replacing in the auxiliary
formulation $-N_f\rho/g^2>0$ by $-N_fU_{aux}(\rho)$ and $g^2/(2N_f)I$ of the defining formulation
by $1/N_fU_{GN}(I)$. Then the saddle point equation is
$\langle(\bar\psi\psi)\rangle=\sigma U'_{aux}(\rho)$, and the matching of the
$N_f=\infty$ potentials leads to the mean-field relation
\be
U_{GN}(I)=N_f^2(2\rho U'_{aux}(\rho)-U_{aux}(\rho))
\ee
This relation implies that a stable power-like asymptotic behavior
$-U_{aux}\sim a_n\rho^n,~~a_n>0$ corresponds to an asymptotic behavior
$-(2n-1)a_nI^n$ for $U_{GN}$. It is natural to expect that the established characteristics of $U(I)$ is carried over also to the finite $N_f$ case.

The derivation of the Wetterich equation for the effective action with $x$-dependent Fermi-fields ($\bar\psi(x),\psi(x)$) and their transposed doublers $(\psi^T(x),\bar\psi^T(x)$) closely follows the steps presented in our previous publication \cite{jakovac13}. We start with a form where the traces with respect the bispinor and flavor indices have been already done:
\be
\partial_k\Gamma_k=-\frac{1}{2}\hat\partial_k{\textrm {Tr}_x}\left[\log G_k^{-1}+\log G_k^{(T)-1}-\log\left(1+(\bar\psi G_k\tilde U\psi)+(\psi^TG_k^{(T)}\tilde U\bar\psi^T)\right)\right].
\label{wetterich-GN-explicit}
\ee
Here $(\bar\psi G_k\tilde U\psi)$ stands for
$\bar\psi(x)^\alpha_jG_{k}^{\alpha\beta,jl}(x,y)\tilde
U(y)\psi^\beta_l(y)$ and summation is understood over all discrete
indices. Similar detailed expression corresponds to
$(\psi^TG_k^{(T)}\tilde U\bar\psi^T)$. The inverse of $G_k,
G_k^{(T)}$, the flavor-diagonal, infrared regularized propagators are
given as
\be
G_k^{-1}(x,y)= g(x,y)^{-1}\delta_{l_1l_2},\qquad
G_k^{(T)-1}(x,y)= g^{(T)-1}(x,y)\delta_{l_1l_2}
\ee
and 
\be
g^{-1}(x,y)=Z_kF_k[\gamma_m\partial_m](x,y)+m_\psi(x)\delta(x-y),\qquad 
g^{(T)-1}(x,y)=Z_kF_k[\gamma_m^T\partial_m](x,y)-m_\psi(x)\delta(x-y).
\ee
Here $F_k(\gamma_m\partial_m)$ is a non-local functional built with $\gamma_m\partial_m$ which freezes efficiently out the propagation modes with wave numbers below the actual normalization scale $k$. For its Fourier-transform there are several propositions which will appear explicitly below. In these expressions one also introduces
\be
m_\psi(x)=2U'(I(x))(\bar\psi(x)\psi(x)),\qquad\tilde U(x)=2U'(I(x))+4IU''(I(x)).
\ee
Below when discussing the scale dependence of $Z_k$, we  shall use the short hand notation
\be
Q(x,y)=(\bar\psi(x) G_k(x,y)\tilde U(y)\psi(y))+(\psi^T(x)G_k^{(T)}(x,y)\tilde U(y)\bar\psi^T(y)).
\ee

\section{The local potential approximation and its fixed points}

One projects the Wetterich equation on the local potential by
substituting into its right hand side constant Fermi fields
($\bar\psi_0,\psi_0$). After performing the operations indicated on
the right hand side of Eq.(\ref{wetterich-GN-explicit}) one finds an
expression which depends only on the scalar invariant
$(\bar\psi_0\psi_0)$ and was given explicitly with optimized infrared
regulator \cite{litim01} in Eq.(31) of \cite{jakovac13}. This equation
is transformed in two steps into a form convenient for finding the
fixed points and the scaling exponents characterizing the behavior of
different operators in its neighborhood. First, one introduces the
following dimensionless rescaled variables taking into account the
anomalous scaling of the wave function renormalization parameter $ \ln
Z_k\sim -\eta\ln k$:
\be
\overline{I}=k^{2(1-d-\eta)}I,\qquad \overline{U}=k^{-d}U(I)|_{I=k^{-2(1-d)+2\eta} \overline {I}}.
\ee
We search for a non-Gaussian fixed point solution of this equation in
the LPA, where one sets $\eta=0$. In order to make our treatment
easier to follow, we consider a second rescaling related to the
large-$N_f$ scaling of the different quantities: 
\be
x=(4Q_dN_f)^{-2}\overline{I},
\qquad y_k=(4Q_dN_f)^{-1}\overline {U}_k,
\ee 
where $Q_d=S_d/(d(2\pi)^d)$ contains the surface $S_d$ of the $d$-dimensional unit sphere. 
These two steps lead to the following evolution equation for  $y_k(x)$:
\be
\partial_ty_k(x)=-dy_k+2(d-1)xy'_k-\left(1+\frac{1}{4N_f}\right)\frac{1}{1+4y_k^{'2}(x)x}+\frac{1}{4N_f}\frac{1}{1+12y^{'2}_k(x)x+16y'_k(x)y''_k(x)x^2},
\label{rescaled-wetterich}
\ee
where the prime denotes the derivative with respect to $x$ and $\partial_t=\partial/\partial (\ln k)$.

Since these are the coefficients of the Taylor-expansion of the fermionic potential which have physical significance, providing the pointlike limit of the $2n$-fermion vertices, it is adequate to search for the fixed point of this RG-equation in form of a power series:
\be
y_*(x)=\sum_{n=1}^{n_{max}} \frac{1}{n}l_{n*}x^n.
\ee
One finds the following equation for $l_{n*}$:
\be
0=\left[-\frac{d}{n}+2(d-1)\right]l_{n*}+8\left(1-\frac{n}{2N_f}\right)l_{1*}l_{n*}+F[l_{1*},...,l_{n-1*}], \qquad n>1,
\label{taylor-solution}
\ee
where the function $F$ is a nonlinear expression of the coefficients with indices lower than $n$. It is easy to solve it after one finds the non-zero solution of the equation for 
 $l_{1*}$:
\be
0=(d-2)l_{1*}+\left(1-\frac{1}{2N_f}\right)4l^2_{1*}.
\ee
This equation was already given in \cite{jakovac13} for the
$N_f=\infty$ case and was shown to coincide with the result of
\cite{braun11} obtained with an Ansatz truncated at $n=1$. It is
worthwhile to emphasize, that the non-Gaussian fixed point exists in
the physical range of $l_{1*}$ only for $N_f>1/2$. The apparent
singularities at $N_f=1/2$ inherited from the denominator of $l_{1*}$
in later formulae do not have any physical meaning. Using the
value of $l_{1*}$ we can determine higher order coefficients,
too. $l_{1*}=0$ yields $l_{i*}=0$ (the Gaussian fixed point), while in the
non-Gaussian fixed point the $l_{i*}$ coefficients have nontrivial
value.  The explicit expression for $l_{2*}$, for instance, reads
\be
l_{2*}=\frac{(d-2)^4N_f^3(N_f-2)}{(1-2N_f)^3(12-5d-2N_f(4-d))}.
\label{coefficient-l2}
\ee

The evolution of the non-Gaussian fixed point with changing
dimensionality was analytically determined at $N_f=\infty$ in
\cite{braun11}. It was found that the fixed point coordinates of the
couplings $\lambda_{2n}^*$ were all proportional to $d-4$, therefore
the non-Gaussian fixed point merges with the Gaussian one when
$d\rightarrow 4$.
When taking Eq.(\ref{taylor-solution}) at $N_f=\infty$ one finds the following equation
for $l_{2*}^\infty$ ($l_{1*}^\infty=-(d-2)/4$):
\be
0=\left(2-\frac{d}{2}\right)l_{2*}^\infty+F(l_{1*}^\infty).
\ee
This would require in $d=4$ $F(l_{1*}^\infty)=0$, which is not
fulfilled. Therefore there is no non-Gaussian solution for $d=4$ at
$N_f=\infty$ neither in our treatment. 

Keeping, however, the $\sim 1/N_f$ term in (\ref{taylor-solution}) for finite
$N_f$ a non-Gaussian fixed point with finite "coordinates" persists up to $d\leq (8N_f-12)/(2N_f-5)$  as one can see from the denominator of (\ref{coefficient-l2}).  For
finite $N_f$ the existence of an upper critical dimension was not discussed in \cite{braun11}. Our fermionic LPA, analytic for arbitrary values of $N_f$  is not sufficient to settle this question. The
momentum dependence of the 4-Fermi coupling probably strongly
influences the conclusion.

Going to the
lower critical dimension one might remark that for $d\rightarrow 2$
the non-Gaussian fixed point merges with the Gaussian one in agreement with \cite{braun11}. 

\begin{figure}[htbp]
  \centering
  \includegraphics[keepaspectratio,width=0.7\textwidth,angle=0]{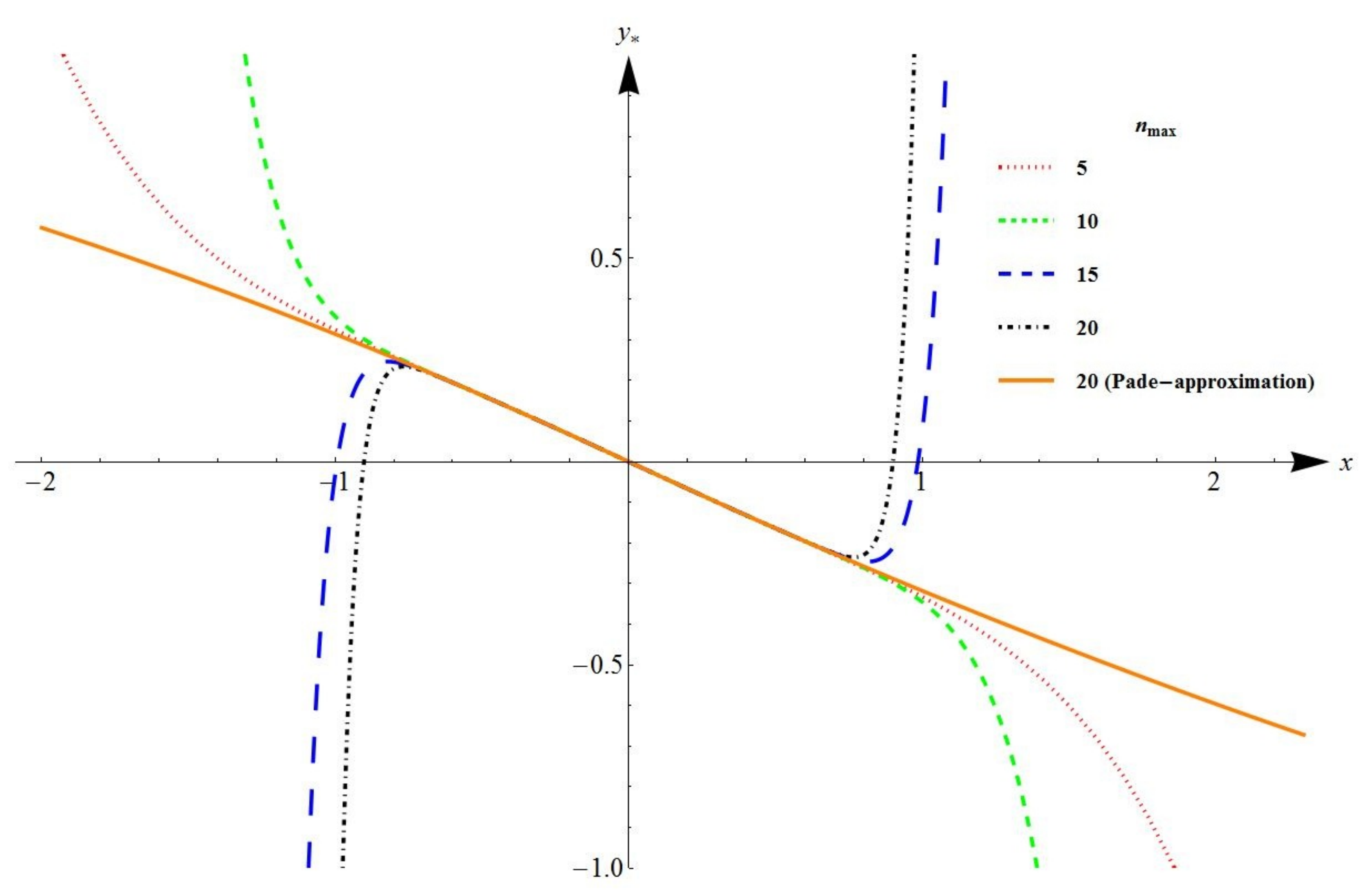}
  \caption{Polynomial approximation of the fixed point potential for
    various $n_{max}$ maximal powers in d=3.}
  \label{fig:nf2abra}
\end{figure}

One can reconstruct from the power series also the potential using the following procedure. The first two terms of the right hand side of
Eq.(\ref{rescaled-wetterich}) determine the large-$x$ asymptotics of
$y_*$: $y_{*as}\sim x^{d/(2(d-1))}$. In $d=3$ the power equals to
$3/4$ and one quickly can check the last two terms on the right
hand side of (\ref{rescaled-wetterich}) asymptotically vanish
consistently. The Taylor series should sum asymptotically into
this power law. 
Being a polynomial, however, they can not converge uniformly to
$x^{d/(2(d-1))}$. This results in a wild oscillation of the different
terms observed also in Ref.~\cite{braun11}.  This behavior is
illustrated in Fig.~\ref{fig:nf2abra}, where the power series of the
potential are presented as obtained with $n_{max}=5,10,15,20$ for
$N_f=2, d=3$, respectively.
One notes the uniform behavior of the series in a finite
and symmetric neighborhood of the origin. This problem is cured by
factoring out $(1+x^2)^{d/(4(d-1))}$ from the power series of the
potential, which is insensitive to the sign of $x$. This factor has
the correct asymptotics, while behaves polynomially for small $x$
values. We know that the ratio of the original power series and of the
asymptotic factor necessarily approaches a constant for large
$x$. This can be achieved using symmetric Pad\'e-approximants to this
ratio. Finally we obtain the fixed point potential with the following
expression:
\be
y_*(x)=(1+x^2)^{d/(4(d-1))}\lim_{N\rightarrow\infty}{\textrm{Pad\'e}}^N_N\left[\frac{\sum_{n=1}^{2N}l_{n*}x^n}{(1+x^2)^{d/(4(d-1))}}\right].
\ee
Here the function ${\textrm{Pad\'e}}^N_N$ refers to the $(N,N)$ Pad\'e
approximant generated from the 2N-th order Taylor series of the
expression in the squared bracket. 

\begin{figure}[htbp]
  \centering
  \includegraphics[keepaspectratio,width=0.7\textwidth,angle=0]{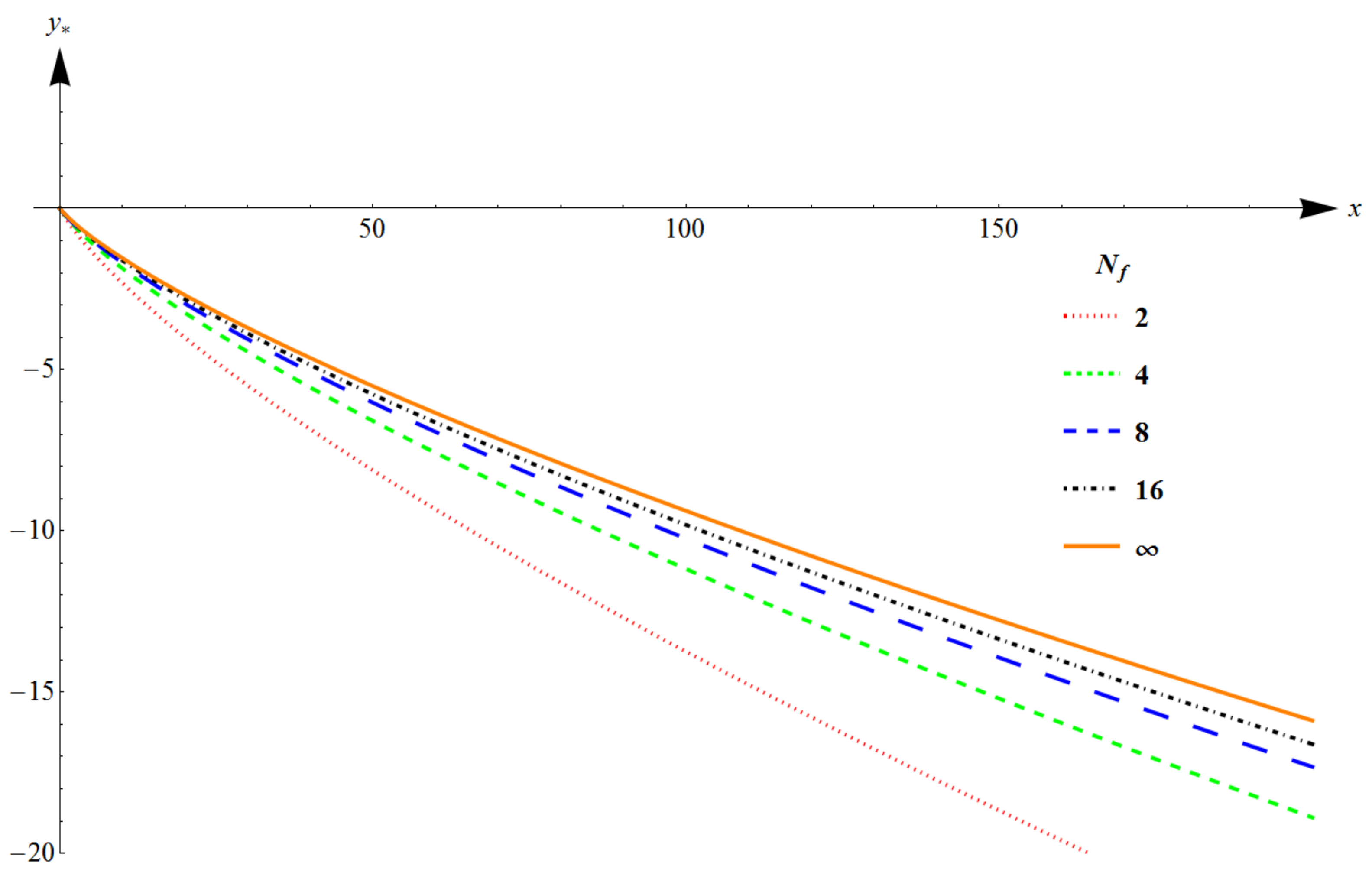}  
  \caption{$N_f$ dependence of the Pad\'e-improved fixed point potentials in $d=3$.}
  \label{fig:padek}
\end{figure}

Using Pad\'e approximation may be dangerous since the polynomials in
its numerator and denominator may produce artificial zeroes and poles,
respectively. However, after separating the correct asymptotics as
described above, we gained the experience that most choices for $N >
12$ led to a smooth and uniformly converging sequence of
potentials. The variation of the fixed point potential with $N_f$ is
illustrated in Fig.~\ref{fig:padek}, where the Pad\'e approximants are
displayed for various $N_f$ values together with the exact numerical
solution of the $N_f=\infty$ case, again for $d=3$.

The flow equation emerging from right hand side of
(\ref{rescaled-wetterich}) for $N_f=\infty$ can be solved with the
method of characteristics \cite{attanasio97,litim11}. Its simple
numerical implementation for the one-variable problem of the fixed
point potential provides a test for the reliability of the method
sketched above. The relative difference between the exact fixed point
potential and the Pad\'e-approximants starts as negligible near the
origin and saturates for $x>10$ below 4\%. In Fig.~\ref{fig:nf2abra}
we also display the resulting fixed point potential together with the
power series approximants. One recognizes the asymptotic series nature
of the Taylor series which coincides on shorter and shorter interval
with the exact solution.

Still, one can make use of this expansion in solving the linearized
eigenvalue problem which by the structure of (\ref{taylor-solution})
is of a lower triangle matrix form. Each power represents therefore an
eigendirection with the following scaling exponents for the
corresponding couplings:
\be
\Theta^{GN}=d-2n\left(1+\frac{(n-1)(d-2)}{2N_f-1}\right).
\label{scaling-exp-GN}
\ee
In $d=3$ there is a single relevant direction ($n=1,
\Theta^{GN}_1=1$), irrespective the value of $N_f$. This result is
compatible with the $N_f=8$ Monte-Carlo simulation of
\cite{karkkainen94}, but deviates from the behavior of the short
series of $1/N_f$-expansion \cite{karkkainen94,hands93}, which display
increasing $N_f$-dependence below $N_f\sim 20$. The exponent
calculated with numerical solutions of the Wetterich-equation as
applied to the auxiliary field formulation stays rather close to the
unit value  for $N_f=3,4$ \cite{rosa01,braun11}. The relevant
eigenvalue of the $N_f=1$ case is significantly
different \cite{hofling02}.

The present spectra of exponents in the $N_f\rightarrow\infty$ limit exactly
reproduces  the result obtained in the partially bosonized
representation \cite{braun11}. The extra Yukawa-coupling $h_k$ is not
running there ($\partial_t h_k=0$) due to the non-trivial scaling
exponent of the auxiliary field. For the remaining (irrelevant) operator set one can establish a clear
correspondence by the scaling exponents between the $2n$-fermion
couplings $l_n$ of our treatment and the $n$-boson vertices
$\lambda_{2n}$ of the auxiliary field formulation. The values found
from (\ref{scaling-exp-GN}) approach the limiting $N_f=\infty$ values
more steeply than the corresponding exponents determined in the
auxiliary field formulation \cite{braun11}. By the correspondence the
renormalization flow pattern in the coupling space $l_n$ is easily
mapped onto the flow in the $\lambda_{2n}$-space of Ref.~\cite{braun11}
around the non-Gaussian fixed point. This correspondence gives some
support to treat $\bar\psi_0\psi_0=\rho$ as a true bosonic field
\cite{aoki14}, but it is cleaner to think in terms of the "average"
correspondence
\be
\frac{1}{\Delta V}\int_{\Delta V} d^dx\bar\psi(x+y)\psi(x+y)\leftrightarrow \rho(y), 
\ee
where $\Delta V$ is the volume defined by the compositeness scale
(cf. the discussion of the fermionic effective potential in the
introduction).

Before proceeding to the investigation of the effects which the
wavefunction renormalization might exert on the above results we
shortly summarize the results of a rather analogous LPA analysis
performed in the $N_f=1$ Nambu--Jona-Lasinio model:
\be
\Gamma^{NJL}_k[\bar\psi,\psi]=\int_x\left[Z_k\bar\psi^\alpha_l(x)\partial_m\gamma_m^{\alpha\beta}\psi_l^\beta(x)+U_k(I_{NJL}(x))\right],\qquad I_{NJL}(x)=\frac{1}{4}\left[(\bar\psi\psi)^2- (\bar\psi(x)\gamma_5\psi(x))^2\right].
\label{Gamma-NJL-ansatz}
\ee
One has to go through the same steps as for the GN-model, starting from
Eq.(51) of Ref.\cite{jakovac13}. The quantities $I_{NJL}$ and
$U^{NJL}$ have the same canonical dimensions like the corresponding
quantities of GN model, furthermore one can scale out also the phase
space factor $Q_d$ rather similarly, one can introduce the scaled
variables:
\be
I_{NJL}=k^{2(d-1)}Q_d^{-2}x,\qquad y_k(x)=Q_d^{-1}k^{-d}U^{NJL}_k(I_{NJL})
\ee
and arrive at the scaled RGE:
\be
\partial_ty_k=-dy_k+2(d-1)xy'_k-\left[\frac{6}{1+xy^{'2}_k}-\frac{1}{1-
xy^{'2}_k}-\frac{1}{1+x(3y^{'2}_k+4y'_ky''_kx)}\right].
\ee

 After introducing the dimensionless
variables, one finds with a Taylor-series search a non-Gaussian fixed
point for the $2n$-fermion couplings. All higher function can be given
in terms of the fixed point value of the $n=1$ coupling 
\be
l_{1*}=-\frac{d-2}{4}.
\ee
The fixed point equation for $l_{n*}$ has the same structure as for
the GN-model:
\be
0=\left[-\frac{d}{n}+2(d-1)-4l_{1*}(n-3)\right]l_{n*}+F^{NJL}\left(l_{1*},...,l_{(n-1)*}\right)
\ee
The corresponding system of linearized flow equations for the
deviation of the point-like limit of the $2n$-fermion couplings from
their fixed point values, denoted as $\delta l_{n}=l_n-l_{n*}$ has
again triangular form 
\be
\partial_t\delta l_n=\left[(2n-1)d-2n-4n(n-3)l_{1*}\right]\delta l_n+n\sum_{j=1}^{n-1}\frac{\partial F^{NJL}(l_1,...,l_{n-1})}{\partial l_j}\Bigr|_{l_*}\delta l_j.
\ee
One finds for the scaling exponents when using the value of $l_{1*}$ :
\be
-\Theta_n^{NJL}=d(n^2-n-1)-2n(n-2).
\ee
The single relevant exponent $\Theta_1=d-2$ is the same as for the Gross-Neveu model, the remaining irrelevant part of the spectra is different.
Here again one can compare our results with earlier investigations of
the pure fermion representation, truncated at $n=1$. The relevant
scaling exponent agrees with the result found for $d=4$ in
\cite{braun12} and also the fixed point value of $l_{1*}$ is the same
if one takes into account our slightly different conventions in
defining spinor variables. As we mentioned before our investigation is
not Fierz-complete, only Fierz-consistent. The study of
Ref.\cite{braun12} was extended to include also vectorial (Thirring-type) interaction
which led also to another non-trivial fixed point, though the number of
relevant operators and the corresponding scaling exponent coincides
with our finding when specified to $d=4$. It has to be noted that for the non-Gaussian fixed point of the NJL-model we did not find any signal for an upper critical dimension.

Although we work with the pure fermionic theory, it is also possible
to estimate the scaling of an indirectly defined effective Yukawa
interaction. The Yukawa interaction ``pulls apart'' the point-like
four-fermion interaction inserting a scalar propagator between pairs
of he external fermion legs, therefore at zero momentum a relation
exists between the coupling constants. Using the bosonic propagator at
zero momentum, it reads $h^2/m_\sigma^2 = \ell_1$, where $h$ is the
Yukawa coupling, $m_\sigma$ is the effective boson mass. For the
scaling of the Yukawa coupling we should take into account that
$\ell_1\sim k^{-\Theta_1},\, m_\sigma^2\sim k^{2-\eta_\sigma}$ (where
$\eta_\sigma$ is the anomalous dimension of the $\sigma$ field). Then
we find $h^2\sim k^{-\Theta_{h^2}}$ with $\Theta_{h^2}
=\Theta_1+\eta_\sigma -2$. This agrees with eq. (40) of
Ref.~\cite{braun11}. To access the bosonic anomalous dimension we need
the bosonic wave function renormalization and use the scaling
$Z_\sigma\sim k^{-\eta_\sigma}$. With the tentative assignment
$\sigma\sim \bar\psi\psi$, the bosonic dynamics should come from the
insertion of the operator
$Z_\sigma[\partial_m(\bar\psi(x)\psi(x))]^2$. This can be a possible
way of extending the fermionic treatment, but in the present
formulation we do not have this operator, thus we have to set zero for
the bosonic anomalous dimension. Therefore now the conjectured scaling
exponent for the Yukawa term is $\Theta_{h^2} = \Theta_1 -2$. Since in
the present model $\eta_\psi=0$ (cf. next section), we find
$\Theta_{h^2}=d-4$.

\section{Wavefunction renormalization in the Gross-Neveu model}

The projection of the Renormalization Group equation on the wave function renormalization constant is given by the equation:
\be
\partial_kZ_k\frac{\delta(0)}{(2\pi)^d}=\frac{1}{N_f}\frac{d}{dq^2}\left\{-iq_m\gamma_m^{\alpha_1\alpha_2}\frac{\delta}{\delta\bar\psi_l^{\alpha_2}(-q)}\partial_k\Gamma_k\frac{\delta}{\delta\psi(q)^{\alpha_1}_l}\right\}_{|q=0}.
\label{wf-renormalisation}
\ee
The task is to substitute into the right hand side of this equation
the three terms on the right hand side of the RGE
(\ref{wetterich-GN-explicit}). Diagrammatically, one has contributions
from the set of Feynman-diagrams illustrated on
Fig.\ref{one-loop-vertex}, just two of the legs are not static. It is
clear that there is a non-trivial dependence on the external momentum
since these diagrams are overwhelmingly not tadpole-type.
 
One observes that the result of the operations prescribed in
(\ref{wf-renormalisation}) still depends on the background
field. Generally one chooses its homogeneous value characterizing the
ground state, that is the minimum of the effective fixed point
potential. This principle dictates in the present case by the global
features of $U(I)$ established in section 2 to choose
$I_0=(\bar\psi_0\psi_0)^2=0$. The experience with various model investigations shows that the anomalous dimension $\eta_k=-\ln Z_k$ is proportional to the
invariant of the theory, and therefore in the symmetric phase $\eta=0$
\cite{berges97}. Still one has to put $I_0=0$ only after carefully
checking that it does not lead in the relevant integrals to infrared
divergences, since the mass term in the propagators is proportional to
this quantity.

In studies of pure fermionic formulation truncated at low powers of
the invariants, the anomalous fermionic wavefunction exponent
$\eta_\psi$ was found to vanish both in the GN- and the NJL-models
\cite{gies10,braun12}. With the non-truncated Ansatz the computation
on the right hand side of (\ref{wf-renormalisation}) becomes quite
tedious. Some of its details are worth to be presented, which follows
below for the GN-model.

We start with ${\textrm {Trlog}G_k^{-1}}$, and promptly use that $G_k^{-1}$ is diagonal in flavor. Its contribution can be expressed as
\bea
&&
\hat\partial_k\frac{i}{2}\frac{d}{dq^2}\int_{y_1}\int_{y_2}e^{iq(y_1-y_2)}\Bigl\{q_m\gamma_m^{\alpha_1\alpha_2}\int_{x_1}\int_{x_2}\Bigl[g(x_1,x_2)
\left(\delta_{\bar\psi_l^{\alpha_2}(y_2)}g^{-1}(x_2,x_1)\delta_{\psi_l^{\alpha_1}(y_1)}\right)
\nonumber\\
&&
-\int_{x_3}\int_{x_4}\left(\delta_{\bar\psi_l^{\alpha_2}(y_2)}g^{-1}(x_2,x_1)\right)g(x_1,x_3)\left(g^{-1}(x_3,x_4)\delta_{\psi_l^{\alpha_1}(y_1)}\right)g(x_4,x_1)\Bigr]\Bigr\}_{|q=0}
\label{g-1-contribution}
\eea
(starting from here we use an abbreviated notation for the functional derivative). Since the regularized kinetic parts of $g^{-1}$ and $g^{(T)-1}$ do not depend on the fermion fields, a straightforward calculation gives for the different terms in the above integrands eventually evaluated on a constant $\psi_0$ background the following expressions:
\bea
\delta_{\bar\psi_{l_1}^{\alpha_2}(y_2)}g^{-1}(x_2,x_1)\delta_{\psi_{l_2}^{\alpha_1}(y_1)}&=&\delta(x_1-x_2)\delta(y_2-x_1)\delta(y_1-x_1)\left[2(\bar\psi_0\psi_0)\tilde U'_0\psi^{\alpha_2}_{0 l_2}\bar\psi^{\alpha_1}_{0 l_1}+\tilde U_0\delta^{\alpha_1\alpha_2}\delta_{l_1l_2}\right],\nonumber\\
\delta_{\bar\psi_{l_2}^{\alpha_2}(y_2)}g^{-1}(x_2,x_1)&=&
\delta(x_1-x_2)\delta(y_2-x_1)\tilde U(I_0)\psi_{0l_2}^{\alpha_2},\nonumber\\
g^{-1}(x_3,x_4)\delta_{\psi_{l_1}^{\alpha_1}(y_1)}&=&\delta(x_3-x_4)\delta(y_1-x_3)\bar\psi^{\alpha_1}_{0l_1}\tilde U(I_0).
\label{psi-derivatives-1}
\eea
It is obvious that when substituting these expressions into the appropriate parts of (\ref{g-1-contribution}) one encounters either $\gamma^{\alpha_1\alpha_2}_m\delta^{\alpha_2\alpha_1}=0$ or $\bar\psi^{\alpha_1}_0\gamma^{\alpha_1\alpha_2}_m\psi_0^{\alpha_2}$. This latter is not included into the Ansatz (\ref{Gamma-ansatz}), therefore we drop it also on the right hand side of the Wetterich equation. The same analysis goes through for ${\textrm{Trlog}}~G_k^{(T)-1}$, therefore even before setting $I_0$ to zero one recognizes that the first two terms of (\ref{wetterich-GN-explicit}) do not contribute to the running of $Z_k$.

For the evaluation of the contribution from the last term of (\ref{wetterich-GN-explicit}) one can write an expression structurally identical to (\ref{g-1-contribution}):
\bea
&&
\hat\partial_k\Biggl(-\frac{i}{2N_f}\frac{d}{dq^2}\int_{y_1}\int_{y_2}e^{iq(y_1-y_2)}q_m\gamma_m^{\alpha_1\alpha_2}\Bigl\{\int_{x_1}\int_{x_2}\Bigl[(1+Q(x_2,x_1))^{-1}
\left(\delta_{\bar\psi_l^{\alpha_2}(y_2)}Q(x_1,x_2)\delta_{\psi_l^{\alpha_1}(y_1)}\right)
\nonumber\\
&&
-\int_{x_3}\int_{x_4}\left(\delta_{\bar\psi_l^{\alpha_2}(y_2)}Q(x_1,x_2)\right)(1+Q(x_2,x_3))^{-1}\left(Q(x_3,x_4)\delta_{\psi_l^{\alpha_1}(y_1)}\right)(1+Q(x_4,x_1))^{-1}\Bigr]\Bigr\}_{|q=0}\Biggr).
\label{psi-derivatives-2}
\eea
After  the tedious but straightforward computation of the derivatives one substitutes the constant spinorial background and exploits that on such background the propagators $g$ and $g^T$ are translationally invariant and also 
\be
Q(x,y)=(\bar\psi_0g_k(x-y)\tilde U(I_0)\psi_0)+(\psi^{T}_0g_k^{(T)}(x-y)\tilde U(I_0)\bar\psi^T_0).
\ee
The Fourier transform of the infrared regularized propagators $g$ and $g^{(T)}$ on a constant $\psi_0$ background read as
\be
g(p)=\frac{-iZp_m\gamma_m(1+r_{kF}(p))+m_\psi}{Z^2P_F(p^2)+m_\psi^2},\qquad
g^{(T)}(p)=\frac{-iZp_m\gamma^T_m(1+r_{kF}(p))-m_\psi}{Z^2P_F(p^2)+m_\psi^2},
\ee
where $r_{kF}(p)$ is the regularizing modification of the kinetic term and $P_F(p^2)=p^2(1+r_{kF}(p))^2$. Throughout the calculation we use the linear regulator \cite{litim01}:
$r_{kF}(p)=(k/\sqrt{p^2}-1)\Theta(k^2-p^2)$.
The propagators determine also the Fourier transform of $Q(x-y)$
\be
Q(p)=\frac{4I_0U^{'}_0\tilde U_0}{Z^2P_F(p^2)+4U^{'2}_0I_0},
\ee 
($U_0\equiv U(I_0), U^{'}_0\equiv U^{'}(I_0)$ etc.).

Also here we omit all terms which would be proportional to the vectorial condensate $\bar\psi\gamma_m\psi$ or give zero after the multiplication by $\gamma^{\alpha_1\alpha_2}$ (see above). When expanding the occurring integrals to linear order in the external momentum $q$ one encounters expressions proportional to $\partial P_F(p^2)/\partial p^2=1-\Theta(k^2-p^2)$ or to $\partial r_F(p^2)/\partial p^2=-k/(2p^3)\Theta(k^2-p^2)$. The coefficients of all integrals are proportional to some power of $I_0$.

The presence of $\partial P_F(p^2)/\partial p^2$ excludes the infrared region from the integration.  In the integrals where the integrand is proportional to $\partial r_F(p^2)/\partial p^2$ the $p^2$-dependence of the rest of the integrands comes from the infrared regularized propagators. These terms, however in the infrared region are frozen to constants, therefore the infrared contribution to the integral is usually of the form: $\int dp p^{d-1}(pq)^2/p^3$. It is regular for $d>2$. Therefore in both types of integrals one can safely send the coefficients to zero.

Finally, there are also contributing integrals which are of the general form
\be
\int_p\frac{(pq)^2(1+r_F(p))^l}{(Z^2P_F(p^2)+m_\psi^2)^k}.
\ee
Since the term $1+r_F(p)\sim p^{-1}$ in the infrared region where the propagators are $p$-independent, these integrals are infrared regular for $d+1>l$, what is true for all occurring cases. Again, one is allowed to set in the coefficients of these integrals $I_0=0$.

The whole rather tiresome discussion of some 15 integrals leads to the short conclusion that in the present formulation of the fermionic FRG:
\be
\partial_tZ_k=0,\qquad {\textrm{all}}~~N_f.
\ee

\section{Discussion}

The existence of a non-Gaussian fixed point in the Gross-Neveu model in $d=3$ with the single relevant (infrared repulsive) operator $I=(\bar\psi\psi)^2$ around which  theories with infinite number of fermion flavors ($N_f=\infty$) and with four-fermion coupling can be consistently renormalized has been established quite some time ago by investigating the ultraviolet behavior of the four-point function of the theory both with \cite{rosenstein89} and without \cite{gawedzki85,calan91} introducing an auxiliary field $\sigma(x)\sim\bar\psi(x)\psi(x)$.  

The present investigation confirmed the UV-safe behavior of the
three-dimensional $(d=3)$ four-fermion models relying on the analysis
of the functional renormalization group equations derived without
introducing any auxiliary Bose-fields. The spectra of scaling
exponents appearing for the Gross-Neveu model in (\ref{scaling-exp-GN}) for $N_f=\infty$ fully
reproduces the exponents found in \cite{braun11} using the auxiliary
formulation of the model.  In case of this
latter approach the running of the auxiliary field renormalization
($\eta_\sigma=4-d$) is essential.  Our finding of zero anomalous dimension of the
fermions even for finite $N_f$ is in qualitative agreement with the rather small $\eta_\psi$ found in the auxiliary field
reformulation \cite{braun11}. 

The existence of a non-Gaussian fixed point with a single infrared
unstable direction appears for $d>2$. Another important issue is to
see if there is an upper dimension $d_{max}$ for the existence of the
non-Gaussian fixed point. According to the analytic solution of the
auxiliary field formulation of the ERG equations for $N_f=\infty$
\cite{braun11} the Gaussian and the non-Gaussian fixed points merge in
$d=4$. In our approach at $N_f=\infty$ the fixed point value of the
4-fermion coupling ($l_{2*}$) is pushed to infinity, which confirms
that there is no consistent non-Gaussian fixed point in $d=4$. For
finite $N_f$ we find an $N_f$-dependent value for $d_{max}>4$, clearly
indicating the necessity to include momentum dependence into the
effective action already on the 4-fermion level.

In $d=3$ we confirm the existence of a non-Gaussian fixed point with
just one relevant eigendirection in the coupling space for all values
of $N_f=1,2,...$. The spectrum we find slightly deviates from those
established numerically in \cite{braun11}. The non-trivial influence
from the anomalous dimensions of the Fermi-field and in particular, of
the auxiliary field could be the reason for this difference. In
particular by introducing a kinetic term, which intuitively
corresponds to the kinetic term of the auxiliary scalar field
$Z_\sigma[\partial_m(\bar\psi(x)\psi(x))]^2$, also in the present
formulation effects related to the anomalous dimension of the
composites might show up. These effects eventually should result in an
improvement of the spectra of scaling exponents and the estimate for
$d_{max}$. By an intuitive correspondence it also provides an estimate for the scale dependence of the squared Yukawa-coupling of the auxiliary field formulation.

 The analogous results obtained for the Nambu--Jona-Lasinio model suggests that the existence of non-Gaussian fixed point(s) in $d=3$ could be a generic feature of models with 4-fermion invariants.

In conclusion, we find rather encouraging the level of agreement we found in analyzing the fixed point structure of the two model systems with and without bosonic composite fields. This fact hints at prospective efficient usage of the technique developed in \cite{jakovac13} without the introduction of any scalar auxiliary field also in more complicated models.

\section*{Acknowledgments}
This research was supported by the grant K-104292 from the Hungarian Research Fund. We thank D. Litim and P. Mati for valuable information on the application of the method of characteristics.

\end{document}